\documentclass[12pt,a4paper]{article}

\usepackage[utf8]{inputenc}
\usepackage[T1]{fontenc}
\usepackage{listings }
\usepackage{xspace}
\usepackage{setspace}

\usepackage{amsmath}
\usepackage{amsfonts}
\usepackage{amssymb}
\usepackage{color}
\usepackage[english]{babel}
\usepackage{amsthm}
\usepackage{appendix}
\usepackage{enumitem}
\usepackage{hyperref}
\usepackage{rotating}

\usepackage{graphicx,pict2e}

\usepackage{endfloat}

\newcommand*{\eg}{e.g.\@\xspace}
\newcommand*{\ie}{i.e.\@\xspace}

\makeatletter
\newcommand*{\etc}{%
	\@ifnextchar{.}%
	{etc}%
	{etc.\@\xspace}%
}
\newcommand*{\myetal}{%
	\@ifnextchar{.}%
	{\textit{et~al}}%
	{\textit{et~al.}\@\xspace}%
}

\newcommand{\EE}{\mathbb{E}}
\newcommand{\PP}{\mathbb{P}}
\newcommand\independent{\protect\mathpalette{\protect\independenT}{\perp}}
\def\independenT#1#2{\mathrel{\rlap{$#1#2$}\mkern2mu{#1#2}}}

\title{Closed-form variance	estimators for weighted and stratified dose-response function estimators using generalized propensity score}
\date{\today}
\author{Val\'erie Gar\`es\thanks{Univ Rennes, INSA, CNRS, IRMAR - UMR 6625, F-35000 Rennes, France, valerie.gares@insa-rennes.fr} \\
Guillaume Chauvet\thanks{Univ Rennes, ENSAI, CNRS, IRMAR - UMR 6625, F-35000 Rennes, France, chauvet@ensai.fr} \\
David Hajage\thanks{Sorbonne Universit\'e, INSERM, Institut Pierre Louis d'Epid\'emiologie et de Sant\'e Publique, AP-HP, Hôpital Pitié-Salpêtrière, D\'epartement de Sant\'e Publique, Centre de Pharmaco\'epid\'emiologie, Paris, France}
}

\begin{document}

%\title{Closed-form variance	estimators for weighted and stratified dose-response function estimators using generalized propensity score}
%
%\author[1]{Val\'erie Gar\`es}
%
%\author[2]{Guillaume Chauvet}
%
%\author[3]{David Hajage*}
%
%\authormark{VALERIE GARES \textsc{et al}}
%
%
%\address[1]{Univ Rennes, INSA, CNRS, IRMAR - UMR 6625, F-35000 Rennes, France, valerie.gares@insa-rennes.fr}
%
%\address[2]{Univ Rennes, ENSAI, CNRS, IRMAR - UMR 6625, F-35000 Rennes, France, chauvet@ensai.fr}
%
%\address[3]{Sorbonne Universit\'e, INSERM, Institut Pierre Louis d'Epid\'emiologie et de Sant\'e Publique, AP-HP, Hôpital Pitié-Salpêtrière, %D\'epartement de Sant\'e Publique, Centre de Pharmaco\'epid\'emiologie, Paris, France}
%
%\corres{*David Hajage, Sorbonne Universit\'e, INSERM, Institut Pierre Louis d'Epid\'emiologie et de Sant\'e Publique, AP-HP, Hôpital %Pitié-Salpêtrière, D\'epartement de Sant\'e Publique, Centre de Pharmaco\'epid\'emiologie, Paris, France. \email{david.hajage@aphp.fr}}
%
%%\presentaddress{This is sample for present address text this is sample for present address text}

\maketitle

\begin{abstract}
Propensity score  methods are widely used in observational studies for evaluating marginal treatment effects. The generalized propensity score (GPS) is an extension of the propensity score framework, historically developed in the case of binary exposures, for use with quantitative or continuous exposures. In this paper, we proposed variance estimators for treatment effect estimators on continuous outcomes. Dose-response functions (DRF) were estimated through weighting on the inverse of the GPS, or using stratification.  Variance estimators were evaluated using Monte Carlo simulations. Despite the use of stabilized weights, the variability of the weighted estimator of the DRF was particularly high, and none of the variance estimators (a bootstrap-based estimator, a closed-form estimator especially developped to take into account the estimation  step of the GPS, and a sandwich estimator) were able to adequately capture this variability, resulting in coverages below to the nominal value, particularly when the proportion of the variation in the quantitative exposure explained by the covariates was large. The stratified estimator was more stable, and variance estimators (a bootstrap-based estimator, a pooled linearized estimator, and a pooled model-based estimator) more efficient at capturing the empirical variability of the parameters of the DRF. The pooled variance estimators tended to overestimate the variance, whereas the bootstrap estimator, which intrinsically takes into account the estimation step of the GPS, resulted in correct variance estimations and coverage rates. These methods were applied to a real data set with the aim of assessing the effect of maternal body mass index on newborn birth weight.
\end{abstract}

%\keywords{}
%
%\jnlcitation{\cname{%
%\author{Val\'erie Gar\`es},
%\author{Guillaume Chauvet}, and
%\author{David Hajage}} (\cyear{202x}),
%\ctitle{Closed-form variance estimators for weighted and stratified dose-response function estimators using generalized propensity score}, %\cjournal{Stat Med}, \cvol{0000;00:?--?}.}
%
%\maketitle
%
%\footnotetext{\textbf{Abbreviations:} GPS, generalized propensity score; DRF, dose-response function; IPTW, inverse probability of treatment weighting; BMI, body mass index; RMSE, root mean square error; SE, standard error}

\section{Introduction}\label{intro}

In observational cohort studies, confounding may occur when the distribution of baseline covariates differs between treated and control subjects. The propensity score is one of the methods that helps in reducing or minimizing this confounding to get valid inferences on treatment effects. It was first developed for binary or categorical exposures (\cite{ROSENBAUM1983}). In this setting, the propensity score is defined as the probabiblity of being exposed conditionally on baseline characteristics. Different propensity score methods have been proposed to estimate the treatment effects: covariate adjustment using the propensity score (\cite{austin_conditioningpropensityscore_2007}), stratification on the propensity score (\cite{rosenbaum_reducingbiasobservational_1984,lunceford_stratificationweightingpropensity_2004}), propensity-score matching (\cite{austin_methodspropensityscorematching_2009,rubin_matchingusingestimated_1996,abadie_matchingestimatedpropensity_2016}) and propensity score weighting (\cite{rosenbaum_modelbaseddirectadjustment_1987,austin_performancedifferentpropensity_2013,li_weighting_2013,li_balancing_2018}).

In many studies, the exposure of interest is continuous rather than binary. For example, we may not only know whether an individual is a smoker or not, but also the pack-years of cigarettes smoked, or the duration of smoking. Another example is the body mass index, which may be more informative as a continuous variable than if reduced to a dummy variable indicating obesity (\cite{zhang2016causal,austin2018assessing}). Considering this type of exposure variable, one may be interested in estimating the dose-response function. If this term may evoke the dose of a medication, we will use it regardless of the nature of the exposure as long as it is quantitative. The propensity score has been generalized into a propensity function for quantitative exposures which is known as the generalized propensity score (GPS) (\cite{hirano2004propensity,imai2004causal,bia2008stata,zhang2016causal,austin2018assessing}). Similarly to the binary case, different propensity score methods have been proposed to estimate the treatment effects on outcomes using the GPS: covariate adjustment (\cite{austin2018assessing}), stratification (\cite{zhang2016causal}) and inverse probability of treatment weighting (IPTW) (\cite{zhang2016causal,austin2018assessing}).

In the case of binary exposure, several authors have proposed valid closed-form  variance estimators adapted to each treatment effect estimators: ad\-just\-ment (\cite{zou_variance_2016}), stratification (\cite{williamson_variancereductionrandomised_2014}), matching (\cite{abadie_matchingestimatedpropensity_2016}) and weighting (\cite{lunceford_stratificationweightingpropensity_2004, lunceford_stratificationweightingpropensity_2017, hajage2018closed}). Note that all these estimators take into account the fact that the theoretical propensity score value of an individual is unknown, and is estimated from the data in the first-stage of analysis. To our knowledge, variance estimation for treatment effect estimated using the GPS framework has received little attention. In this work, we develop and evaluate closed-form variance estimators for stratified and weighted treatment effect estimators using the influence function linearization technique (\cite{deville_varianceestimationcomplex_1999}). These variance estimators are also compared to bootstrap-based variance estimators.

The paper is organized as follows. In Section \ref{notations}, we introduce some notations. In Section \ref{weight}, we describe the weighted treatment effect estimator based on the GPS. In Section \ref{strat}, we describe the stratified treatment effect estimator. In Section \ref{variance}, we describe the variance estimators developed in this study. In Section \ref{simul}, the performances of the models are assessed on a benchmark of simulated databases. They are applied in Section \ref{example} on a real example extracted from the PreCARE cohort study, with the aim of assessing the effect of maternal Body Mass Index (BMI) on newborn birth weight. Finally, we discuss in Section \ref{conclusion} the pros and cons of the different estimation methods, and we describe areas for future research.

\section{Treatment effect estimator using the generalized propensity score}\label{effect}

\subsection{Notations and assumptions}\label{notations}
Let $T$ denote the level of a quantitative exposure which is a continuous  variable, and  $Z$  a set of $p$ baseline measured covariates. Let $Y(t),~t \in \Psi$, denote a set of potential outcomes which is assumed to exist under  Rubin's framework for causal inference. More precisely, we assume that  $T$ is a continuous exposure (\ie, $\Psi$ is a subset of $\mathbb{R}$) and that $Y_i(t)$ is the outcome that would be observed for subject $i$ if he/she received (maybe contrary to the reality) the level of exposure $T = t$. In practice, we only observe one level of exposure for each subject $i$ and the corresponding outcome. The observed data consists of $(Z_i, T_i, Y_i)$ for subjects $i=1,\ldots,n$.\\

We are interested in estimating the dose-response function
\begin{eqnarray} \label{muT}
\mu(t) & = & \EE[Y_i(t)],
\end{eqnarray}
which corresponds to the average response if all subjects were exposed to the level $T = t$.

In randomized studies, it can be assumed that $Y(t)$ is independent of $T$, which is denoted as $Y(t) \independent T , \quad \forall  t \in \Psi$. In this work, we only assume that $Y(t)$ is independent of $T$ given $Z$, which is known as the weak unconfoundedness assumption and denoted as
\begin{eqnarray}  \label{weak:ass}
Y(t) \independent T | Z, \quad \forall  t \in \Psi.
\end{eqnarray}
This assumption means that any association between the actual exposure and the potential outcomes is explained by  a set of baseline covariates $Z$ (\cite{hirano2004propensity,zhang2016causal}). Note that this assumption cannot be checked from the data. \\

Let us denote by
\begin{eqnarray} \label{r:tz}
r(t \mid z) & \equiv & f_{T|Z}(t|z),
\end{eqnarray}
the conditional density of exposure variable $T$ given the covariates, which is called the generalized propensity score (GPS) (\cite{hirano2004propensity}). We make the positivity assumption, namely
\begin{eqnarray} \label{pos:ass}
r(t \mid z)>0 & \textrm{ for any } & t \in \Psi \textrm{ and for any } z.
\end{eqnarray}
This means that any level of exposure $T=t$ is possible for any subject, whatever his/her baseline characteristics. A violation of this assumption may lead to biased estimators, or estimators with a large variability (\cite{moore2012causal}). Note that this assumption may and should be checked from the data. In the case of a binary exposure, assessing the positivity assumption may involve examining the overlap between the distribution of the estimated propensity score for the exposed and the exposed samples (\cite{mccaffrey_tutorial_2013}), or by examining the distribution of the estimated weights used for inverse probability of treatment weighting, looking for extreme values (\cite{austin_moving_2015}). In the context of the generalized propensity score and to our knowledge, diagnostics for assessing the positivity assumption have not yet received much attention, even though the estimation of the proportion of the variation in the continuous exposure explained by the covariates seems a promising approach (\cite{austin2018assessing}).

We focus on two approaches for the estimation of the dose-response function: inverse probability of treatment weighting, and stratification. Both approaches are presented in \cite{zhang2016causal} and are briefly described in Sections \ref{weight} and \ref{strat}.

\subsection{Weighted treatment effect estimator} \label{weight}

The first estimator is obtained by fitting a generalized linear regression model between the dose-response function and the exposure, used as the sole dependent variable. Focusing on the case of a linear dose-response function, our model is
\begin{eqnarray} \label{mod1}
\mu(T) & = & \beta_0 + \beta_1~T + \epsilon,
\end{eqnarray}
where $\beta_0$ is the average response observed in case of null exposure ($T = 0$), and $\beta_1$ is the average response change if the level of exposure is increased by one unit. Other dose-response functions (\eg in case of non-linear relationship) and/or other link functions may be better suited for other types of outcome (\eg, a binomial link function for a binary outcome), and may therefore be alternatively used. \\

The parameter $\beta=(\beta_0,\beta_1)^{\top}$ is estimated by weighted least squares, which leads to
\begin{eqnarray} \label{betahat:w}
\hat{\beta}_{w} & \equiv & (\hat{\beta}_{w0},\hat{\beta}_{w1})^{\top} = \left(\sum_{i=1}^n \hat{w}_i \tilde{T}_i \tilde{T}_i^{\top} \right)^{-1} \left(\sum_{i=1}^n \hat{w}_i \tilde{T}_i Y_i \right)
\end{eqnarray}
where $\tilde{T}_i=(1,T_i)^{\top}$, and the weights $\hat{w}_i$ that we use are presented thereafter. This leads to the first estimator
\begin{eqnarray} \label{est:1}
\hat{\mu}_{w}(t) & = & \hat{\beta}_{w0} + \hat{\beta}_{w1}~t
\end{eqnarray}
for the dose-response function. \\

The weights used in equation (\ref{betahat:w}) are computed as follows. We first introduce the theoretical Generalized Propensity Score (GPS) weights, defined as
\begin{eqnarray} \label{gps:weight}
w_i(\gamma) & = & \frac{W(T_i|\gamma)}{r(T_i|Z_i,\gamma)},
\end{eqnarray}
where $r(t|z,\gamma)$ is the conditional density of the exposure variable defined in (\ref{r:tz}), and where $W(\cdot|\gamma)$ is a stabilization factor. As is currently done in the literature, we use $W(t|\gamma) \equiv f_T(t|\gamma)$ the marginal density of the exposure variable. Note that the weights depend on some unknown vector of parameters $\gamma$, which needs to be estimated. \\

We suppose that $T_i$ follows a normal distribution, both conditionally on $Z_i$ and non conditionally. We may therefore write
\begin{eqnarray}
f_T(t|\gamma) & = & \frac{1}{\sqrt{2\pi \sigma_T^2}} \exp\left\{-\frac{1}{2 \sigma_T^2} \left(T_i-\mu_T\right)^2 \right\}, \label{f_T:norm} \\
r(t|z,\gamma) & = & \frac{1}{\sqrt{2\pi \sigma^2}} \exp\left\{-\frac{1}{2 \sigma^2} \left(T_i-\alpha^{\top} \tilde{Z} _i  \right)^2 \right\}, \label{r_T:norm}
\end{eqnarray}
with $\tilde{Z}_i^{\top}=(1,Z_i^{\top})$ and $\gamma=(\mu_T,\sigma_T^2,\alpha^{\top},\sigma^2)^{\top} $. The parameters $\mu_T$ and $\sigma_T^2$ in equation (\ref{f_T:norm}) are estimated by
\begin{eqnarray} \label{est:param:fT}
\widehat{\mu}_{T}=\frac{1}{n}\sum_{i=1}^n T_i & \textrm{and} & \widehat{\sigma}^2_{T}=\frac{1}{n-1}\sum_{i=1}^n (T_i-\widehat{\mu}_{T})^2.
\end{eqnarray}
By fitting a linear regression model between the exposure variable and the covariates, namely
\begin{eqnarray} \label{prop:mod}
T_i & = & \tilde{Z_i}^{\top} \alpha + \eta_i,
\end{eqnarray}
the parameters $\alpha$ and $\sigma^2$ in equation (\ref{r_T:norm}) are estimated by
\begin{eqnarray} \label{est:param:rT}
\hat{\alpha} & = & \left(\sum_{i=1}^n \tilde{Z}_i \tilde{Z}_i^{\top} \right)^{-1} \left(\sum_{i=1}^n \tilde{Z}_i T_i \right), \\
\hat{\sigma}^2 & = & \frac{1}{n-p-1} \sum_{i=1}^n (T_i-\widehat{\alpha}^T \tilde{Z}_i)^2. \nonumber
\end{eqnarray}
This leads to the estimator $\hat{\gamma}=(\hat{\mu}_T,\hat{\sigma}_T^2,\hat{\alpha}^{\top},\hat{\sigma}^2)^{\top}$. By plugging this estimator in (\ref{gps:weight}), we obtain the estimated weights $\hat{w}_i \equiv w_i(\hat{\gamma})$ used in equation (\ref{betahat:w}). The model (\ref{prop:mod}) is called the propensity model in the remainder of this paper.

\subsection{Stratified treatment effect estimator}\label{strat}

The weighted estimator of the dose-response function considered in equation (\ref{est:1}) of Section \ref{weight} proceeds through  a linear regression on the whole sample, using weights to adjust for possible imbalance in the covariates. \\

An alternative approach consists in partitioning the sample into $L$ strata, in such a way that the units inside a given stratum are  somewhat similar with respect to the covariates. This may be done by fitting the propensity model in (\ref{prop:mod}), ordering the units in the sample with respect to the prediction $Z_i^{\top} \widehat{\alpha}$, and using the quantiles as cut-off points (\cite{zhang2016causal}). \\

Inside any stratum $l=1,\ldots,L$, we fit the regression model
\begin{eqnarray} \label{mod:st}
\mu(T) & = & \beta_{l0} + \beta_{l1}~T + \epsilon_l,
\end{eqnarray}
and by estimating the parameter $\beta_l=(\beta_{l0},\beta_{l1})^{\top}$ by ordinary least squares, we obtain
\begin{eqnarray} \label{betahat:st:1}
\hat{\beta}_{l} & \equiv & (\hat{\beta}_{l0},\hat{\beta}_{l1})^{\top} = \left(\sum_{i \in S_l} \tilde{T}_i \tilde{T}_i^{\top} \right)^{-1} \left(\sum_{i \in S_l} \tilde{T}_i Y_i \right),
\end{eqnarray}
with $S_l$ the subset of sampled units which belong to the stratum $l$. The stratified estimator of the parameter $\beta$ in (\ref{mod1}) is obtained by pooling these $L$ estimators, which leads to
\begin{eqnarray} \label{betahat:st}
\widehat{\beta}_{st} & \equiv & (\hat{\beta}_{st0},\hat{\beta}_{st1})^{\top}
= \sum_{l=1}^L\frac{n_l}{n} \hat{\beta}_{l},
\end{eqnarray}
with $n_l$ the number of sampled units in the stratum $S_l$. Note that if the quantiles are used as cut-off points, we have (up to rounding) $n_l=n/L$, and $\widehat{\beta}_{st}$ is the simple mean of the estimators $\hat{\beta}_{l},~l=1,\ldots,L$. \\

This leads to the second estimator
\begin{eqnarray} \label{est:2}
\hat{\mu}_{st}(t) & = & \hat{\beta}_{st0} + \hat{\beta}_{st1}~t
\end{eqnarray}
for the dose-response function. Again, ordinary least squares may be replaced by a generalized linear model and appropriate link function to fit other types of outcome.

\section{Closed form variance estimators}\label{variance}

In this Section, our objective is to develop closed-form variance estimators for the estimators of the dose-response function presented in equations (\ref{est:1}) and (\ref{est:2}). Without loss of generality, we focus on variance estimation for the estimated coefficients of regression $\hat{\beta}_{w}$ and $\widehat{\beta}_{st}$. \\

We follow the influence function linearization technique developped by Deville (\cite{deville_varianceestimationcomplex_1999}), see also \cite{hajage2018closed}. For an estimator $\hat{\beta}$, this technique consists in finding a so-called estimated linearized variable $\hat{I}_i$, summarizing the variability in the estimation of the parameter. Ideally, the linearized variable should account for all the estimation steps which lead to the estimator $\hat{\beta}$. \\

The proposed variance estimator for the weighted estimator $\hat{\beta}_{w}$ presented in Section \ref{weight} is given in Section \ref{evar:weight}. The proposed variance estimator for the stratified estimator $\hat{\beta}_{st}$ presented in Section \ref{strat} is given in Section \ref{evar:strat}.

\subsection{Weighted treatment effect estimator } \label{evar:weight}

The variance estimator for $\hat{\beta}_{w}$ is obtained by observing that the coefficient of regression is estimated in a two-step process, involving two estimating equations. First, the unknown parameter $\gamma$ used to compute the weights is obtained by solving the system of estimating equations
\begin{eqnarray} \label{Fn:gamma}
F_n (\gamma) \equiv \frac{1}{n} \sum_{i=1}^n F_i(\gamma) & = & 0,
\end{eqnarray}
where
\begin{eqnarray} \label{Fi:gamma}
F_i(\gamma) & = & \left(
\begin{array}{l}
T_i-\mu_T \\
(T_i-\mu_T)^2-\frac{n-1}{n} \sigma_T^2 \\
(T_i-\tilde{Z}_i^{\top} \alpha) \tilde{Z}_i \\
(T_i-\tilde{Z}_i^{\top} \alpha)^2-\frac{n-p-1}{n}\sigma^2
\end{array}
\right).
\end{eqnarray}
Then, the estimator $\hat{\beta}_{w}$ is obtained as the solution of the estimating equation
\begin{eqnarray} \label{ee:beta}
H_n (\hat{\gamma},\beta) \equiv \frac{1}{n} \sum_{i=1}^n w_i(\hat{\gamma}) H_i(\beta) = 0 & \textrm{with} & H_i(\beta) = \tilde{T}_i (Y_i-\tilde{T}_i^{\top} \beta).
\end{eqnarray}

After some algebra, this leads to the following linearized variable for $\hat{\beta}_{w}$:
\begin{eqnarray} \label{lin:weight:1}
\hat{I}_{1i} & = & \hat{A}^{-1} \left\{w_i(\hat{\gamma})H_i(\hat{\beta})+\hat{B} \hat{C}^{-1}F_i(\hat{\gamma}) \right\},
\end{eqnarray}
with
\begin{eqnarray}
\hat{A} & = & \frac{1}{n} \sum_{i=1}^n w_i(\hat{\gamma}) \tilde{T}_i \tilde{T}_i^{\top}, \nonumber \\
\hat{B} & = & \frac{1}{n} \sum_{i=1}^n H_i(\hat{\beta}) \nabla w_i(\hat{\gamma})^{\top}, \label{lin:weight:2} \\
\hat{C} & = & \left(
\begin{array}{cccc}
1 & 0 & 0_{1K} & 0 \\
0 & \frac{n-1}{n} & 0_{1K} & 0 \\
0_{K1} & 0_{K1} & \frac{1}{n} \sum_{i=1}^n \tilde{Z}_i \tilde{Z}_i^{\top} & 0_{K1} \\
0 & 0 & 0 & \frac{n-p-1}{n}
\end{array}
\right), \nonumber
\end{eqnarray}
and where $0_{\bullet \diamond}$ stands for the null matrix with $\bullet$ rows and $\diamond$ columns. The computation details are given in Appendix \ref{appen1}. \\

The resulting variance estimator is
\begin{eqnarray} \label{v:lin:weight}
\hat{V}_{lin}(\hat{\beta}_{w}) = \frac{1}{n(n-1)} \sum_{i=1}^n (\hat{I}_{1i}-\bar{I}_1)(\hat{I}_{1i}-\bar{I}_1)^{\top}
& \textrm{ with } & \bar{I}_1 = \frac{1}{n} \sum_{i=1}^n \hat{I}_{1i}.
\end{eqnarray}

\subsection{Stratified treatment effect estimator} \label{evar:strat}

Inside each stratum $\ell=1,\ldots,L$, the intermediary estimators $\hat{\beta}_l$ are estimated by solving the estimating equations
\begin{eqnarray} \label{strat:est:eq:1}
\Phi(\beta_{\ell}) & \equiv & \frac{1}{n_\ell} \sum_{i \in S_\ell} \Phi(Y_i,T_i,\beta_{\ell}) = 0,
\end{eqnarray}
with
\begin{eqnarray} \label{strat:est:eq:2}
\Phi(Y_i,T_i;\beta_{\ell}) & = & \left(
\begin{array}{l}
Y_i - \tilde{T}_i^{\top} \beta_{\ell} \\
T_i(Y_i - \tilde{T}_i^{\top} \beta_{\ell})
\end{array}
\right).
\end{eqnarray}

After some algebra, the linearized variable of $\hat{\beta}_l$ is
\begin{eqnarray} \label{strat:est:eq:3}
\hat{I}_{2l,i} & = & \frac{1}{\hat{\sigma}_{\ell,T}^2}
\left(
\begin{array}{cc}
\hat{\sigma}_{\ell,T}^2+\hat{m}_{\ell,T}^2 & -\hat{m}_{\ell,T} \\
-\hat{m}_{\ell,T} & 1
\end{array}
\right) \times
\left(
\begin{array}{l}
Y_i - \tilde{T}_i^{\top} \hat{\beta}_{\ell} \\
T_i(Y_i - \tilde{T}_i^{\top} \hat{\beta}_{\ell})
\end{array}
\right),
\end{eqnarray}
where
\begin{eqnarray} \label{strat:est:eq:4}
\hat{m}_{\ell,T} = \frac{1}{n_\ell} \sum_{i \in S_\ell} T_i
& \textrm{and} &
\hat{\sigma}_{\ell,T}^2 = \frac{1}{n_\ell-1} \sum_{i \in S_\ell} (T_i-\hat{m}_{\ell,T})^2.
\end{eqnarray}
The computation details are given in Appendix \ref{appen2}. This leads to the pooled variance estimator
\begin{eqnarray} \label{pool:vest}
\hat{V}(\hat{\beta}_{st}) & = & \sum_{\ell=1}^L
\frac{(p_\ell)^2}{n_\ell(n_\ell-1)} \sum_{i \in S_l} (\hat{I}_{2l,i}-\bar{I}_{2l})(\hat{I}_{2l,i}-\bar{I}_{2l})^{\top} \\
\textrm{with } \bar{I}_{2l} & = & \frac{1}{n_l} \sum_{i \in S_l} \hat{I}_{2l,i}. \nonumber
\end{eqnarray}

Note that the strata are built by using the quantiles of the predicted $\tilde{Z}_i^{\top} \hat{\alpha}$ given by the propensity model, and the strata boundaries are therefore estimated rather than known. This is not accounted for in the variance estimator proposed in equation (\ref{pool:vest}). Taking this estimation into account could possibly be performed by following the approach in \cite{williamson_varianceestimationstratified_2012}, but this would require fully specifying the joint distribution between the outcome, the exposure and the covariates. \\

An advantage of the variance estimator given in (\ref{pool:vest}) is its robustness to the misspecification of the model linking the dose-response function and the exposure. Alternatively, a model-based variance estimator could be derived.

\section{Simulations}\label{simul}
\subsection{Data-generating process}\label{ss-data}
We adapt the method described in \cite{hajage2018closed}.
First, we randomly generate $p+1$ normally distributed covariates $Z_1, \dots, Z_k \dots, Z_K, Z_U$ from the following multivariate normal distribution:

\begin{equation*}
[Z_1, \dots, Z_K, Z_U] \sim \mathcal{N}(0;\Sigma) \textrm{ with } \Sigma =
\begin{pmatrix}
1         & 0         & \dots    & 0         & \sigma_{1} \\
0         & 1         & \dots    & 0         & \sigma_{2} \\
\vdots    & \vdots    & \ddots   & \vdots    & \vdots    \\
0         & 0         & \dots    & 1         & \sigma_{K} \\
\sigma_{1} & \sigma_{2} & \dots    & \sigma_{K} & 1         \\
\end{pmatrix}.
\end{equation*}

Thus, $Z_1, \dots, Z_K$ are mutually independent following a standard normal distribution, but are each correlated to a standard normal variable $Z_U$ through covariance parameters $\sigma_{k}$, $k=1,\ldots,K$.

A covariate $U$ is then computed by applying the following transformation to $Z_U$: $U = F_{Z_U}(u) = \PP(Z_U < u)$ (\ie $U$ is the cumulative distribution function of $Z_U$). By construction, $U$ follows a uniform distribution $\mathcal{U}(0,1)$ which is still correlated to other covariates $Z_1, \dots, Z_L$.\\

The treatment allocation $T$ is drawn from a linear model where:
\begin{equation}\label{eq:T}
T = \alpha_0 + \sum_{k=1}^K \alpha_k Z_k + \eta,
\end{equation}
with $\eta \sim \mathcal{N}(0,\sigma^2)$. The parameter $\sigma^2$ is linked to the coefficient of determination $R^2$ which measures the proportion of the variance (of the exposure) explained by the regression model, and is defined as:
\begin{eqnarray}
R^2&=&\frac{\mathrm{var}(\sum_{k=1}^K \alpha_k Z_k)}{\mathrm{var}(Y)}\\
&=&\frac{\sum_{k=1}^K \alpha_k^2 }{\sum_{k=1}^K \alpha_k^2 +\sigma^2}. \label{link}
\end{eqnarray}

$R^2$ is bounded between 0 and 1. This simple parameter (classic in linear regression) allows to easily control the degree of confounding in the simulated samples (\cite{austin2018assessing}). $R^2$ close to 0 corresponds to weak confounding, $R^2$ close to 1 corresponds to strong confounding.

The continuous outcome is then generated from $U$ as
\begin{equation}\label{eq:Y}
Y = \beta_0 + \beta_1 T + \sigma^2_Y U,
\end{equation}
and therefore $Y \sim \mathcal{N}(\mu,\sigma^2_Y)$ where $\mu= \beta_0 + \beta_1 T$.
%We have $Y=F^{-1}_{\mathcal{N}(\mu,\sigma^2_Y)}(U)$

\begin{figure}[!htb]
	\centerline{\includegraphics[width=4cm]{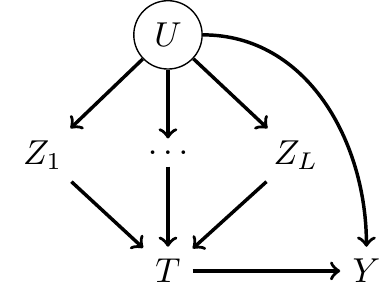}}
	\caption{Directed acyclic graph corresponding to the data-generating algorithm}
	\label{fig:dag}
\end{figure}

The key mechanism by which this algorithm generates confounding in the estimation of the dose response function is the way in which the exposure $T$ and the outcome $Y$ depend  both on $U$. Figure \ref{fig:dag} represent the directed acyclic graph corresponding to  this data-generating process. Confounding is due to $U$ being a common ancestor of $T$ and $Y$. $Z_1, \dots, Z_K$ are sufficient to adjust for confounding, because $T$ is independent of $U$ given $Z_1, \dots, Z_K$ (\cite{havercroft_simulatingmarginalstructural_2012}). Thus, unlike Austin (2018, equation 2, page 1877),\nocite{austin2018assessing} the association between the confounding factors and the outcome is not induced by including these covariates with the exposure $T$ in a conditional equation. By directly setting the vector of parameters $\beta$ of the marginal dose-response function at desired theoretical value, our data generating algorithm allows to evaluate and compare the performance of different analytical methods by their ability to estimate $\beta$ and the variability of this estimation.

\subsection{Simulation parameters}\label{ss-param}
We fixed $K=10$, and the true parameters $\alpha_k$ and $\sigma_k$ were set to values presented in Table \ref{tab:coef} inspired from \cite{austin2018assessing}. Coefficients $\alpha_0$ and $\beta_0$ are fixed to 0 in all scenarios.

Several scenarios were considered, defined by:

\begin{enumerate}
	\item the sample size: $n \in \{500, 1000, 2000\}$;
	\item the degree of confounding tuned by the coefficient of determination $R^2 \in \{0.2,0.4,0.6,0.8\}$
	\item the residual variance in the outcome model $\sigma_Y^2 \in \{0.25,0.5\}$
	\item the treatment effect: $\beta_1  \in \{0,1,2\}$.
\end{enumerate}
A total of $B=1000$ datasets were generated for each scenario.

\begin{center}
	\begin{table}[t]%
		\centering
		\caption{Parameters used in the data-generating process.}\label{tab:coef}
		\begin{tabular*}{200pt}{|cc|cc|}%{@{\extracolsep\fill}llll@{\extracolsep\fill}}
			%\toprule
            \cline{1-4}
			\textbf{Parameter} & \textbf{Value} & \textbf{Parameter} & \textbf{Value}   \\
            \cline{1-4}
			%\midrule
			$\alpha_1$ & $1$ & $\sigma_1$ & $0.2$  \\
			$\alpha_2$ & $1.5$ & $\sigma_2$ & $0.3$  \\
			$\alpha_3$ & $2$ & $\sigma_3$ & $-0.4$  \\
			$\alpha_4$ & $3$ & $\sigma_4$ & $-0.3$  \\
			$\alpha_5$ & $-2$ & $\sigma_5$ & $-0.2$  \\
			$\alpha_6$ & $-2$ & $\sigma_6$ & $0.15$  \\
			$\alpha_7$ & $1$ & $\sigma_7$ & $0.2$  \\
			$\alpha_8$ & $1.5$ & $\sigma_8$ & $-0.2$  \\
			$\alpha_9$ & $2$ & $\sigma_9$ & $-0.2$  \\
			$\alpha_{10}$& $3$ & $\sigma_9$ & $0.2$  \\
            \cline{1-4}
			%\bottomrule
		\end{tabular*}
		%\begin{tablenotes}\centering
		%	\item[$\dagger$] $\alpha_k$: coefficients of the exposure model.
	    %	\item[$\ddagger$] $\sigma_k$: covariance between $Z_k$ covariates and $Z_U$.
		%\end{tablenotes}
	\end{table}
\end{center}

\subsection{Estimation of the parameters of the dose-response function and their variance}\label{ss-methods}

All statistical methods estimating the dose-response function described in Section \ref{effect} were applied to each simulated dataset, and compared to the naive (unweighted) maximum likelihood estimator.

Three different variance estimators were associated with the weighted estimator of the dose-response function. First, we evaluated the sandwich variance estimator  previously used in \cite{zhang2016causal}. This estimator takes into account the lack of independence in the weighted sample (\textit{e.i.} the 'duplication' of subjects in the analysis generated by the weights), but not the fact that the GPS used to derive the weights was estimated rather than known with certainty (\cite{austin_varianceestimationwhen_2016}). The linearized variance estimator proposed in Section \ref{evar:weight} was also applied. Finally, a bootstrap variance estimator based on $N_{\text{boot}} = 200$ bootstrap samples was also used. The weights defined in Equation \ref{gps:weight} and the parameters of the dose-response function ($\hat{\beta}_{0b}$ and $\hat{\beta}_{1b}$, $b \in 1,\dots,N_{\text{boot}}$)  were reestimated in each bootstrap sample. The bootstrap variance estimator was computed as the empirical variance of the estimated regression coefficients associated with dose-response function across the $N_{\text{boot}}$ bootstrap samples.

The stratified estimator of the dose-response function was used with strata defined according to the deciles of the linear predictors of the propensity model. We also considered three variance estimators: the pooled linearized variance estimator given in equation \eqref{pool:vest}, the pooled variance estimator using model-based variance estimator from the maximum likelihood estimator in each stratum, and the boostrap-based variance estimator based on $N_{\text{boot}} = 200$ bootstrap samples. Again, the propensity model and the parameters of the dose-response function were reestimated within each bootstrap samples.

The evaluated methods are summarized in table \ref{tab:methods}.

\begin{center}
\begin{table}[t]
  \caption{Methods for dose-response function and variance estimation}\label{tab:methods}
  \begin{tabular}{ccc}
    \hline
    \textbf{Dose-response function} & \textbf{Variance} & \textbf{GPS estimation} \\
    \textbf{estimator} & \textbf{estimator} & \textbf{taken into account} \\
    & & \textbf{in the variance estimator ?} \\ \cline{1-3}
	Naive & Model-based &  \\ \cline{1-3}
	Weighted & Sandwich &  \\
	Weighted & Linearized & \multicolumn{1}{c}{\checkmark} \\
	Weighted & Bootstrap & \multicolumn{1}{c}{\checkmark} \\ \cline{1-3}
	Stratified & Pooled model-based & \\
	Stratified & Pooled linearized & \\
	Stratified & Bootstrap & \multicolumn{1}{c}{\checkmark} \\ \hline
  \end{tabular}
\end{table}
\end{center}

%	\begin{sidewaystable}[t]%
%		\centering
%		\caption{Different methods for dose-response function and variance estimation}\label{tab:methods}
%		\begin{tabular*}{500pt}{cc}%{@{\extracolsep\fill}l|l@{\extracolsep\fill}}
%			%\toprule
%           \hline
%			\textbf{Dose-response function estimator -- Variance estimator} & \textbf{GPS estimation} \\
%            & \textbf{taken into account} \\
%            & \textbf{in the variance estimator ?} \\
%			\hline
%            %\midrule
%			Naive -- Model-based variance &  \\
%			Weighted -- Sandwich variance &  \\
%			Weighted -- Linearized variance & \multicolumn{1}{c}{\checkmark} \\
%			Weighted -- Bootstrap variance & \multicolumn{1}{c}{\checkmark} \\
%			Stratified -- Pooled model-based variance & \\
%			Stratified -- Pooled linearized variance & \\
%			Stratified -- Bootstrap variance & \multicolumn{1}{c}{\checkmark} \\
%           \hline
%			%\bottomrule
%		\end{tabular*}
%	\end{sidewaystable}
%\end{center}

\subsection{Performance criteria}\label{ss-perf}

Results were assessed in terms of the following criteria:

\begin{enumerate}
	\item Bias of the treatment effect estimation: $\frac{1}{B} \sum_{b=1}^B (\widehat{\beta}_b - \beta)$;
	\item Root mean square error (RMSE): $\sqrt{\frac{1}{B} \sum_{b=1}^B (\widehat{\beta}_b - \beta)^2}$;
	\item Variability ratio of the treatment effect, defined as: 	$\frac{\frac{1}{B}\sum_{b=1}^{B}\widehat{\mathrm{SE}}(\widehat{\beta}_b)}
	{\sqrt{\frac{1}{B^\prime-1}\sum_{b^\prime=1}^{B^\prime}(\widehat{\beta}_{b^\prime}-\bar{\widehat{\beta}})^2}}$,	where $\widehat{\mathrm{SE}}(\widehat{\beta})$ is the estimated standard error of treatment effect $\widehat{\beta}$. It allows evaluating the performance of the variance estimators: a ratio $> 1$ (or $< 1$) suggests that standard errors overestimate (respectively, underestimate) the variability of the estimate of treatment effect. The denominator is the empirical Monte-Carlo standard deviation of the treatment effect, estimated over $B^\prime= 10000$ random samples independent from the samples used in the numerator.
	\item Finally, the coverage evaluates if the procedure for constructing the 95\% confidence interval achieves the advertised nominal level.  Ninety-five percent confidence intervals were constructed by as $\widehat{\beta} \pm 1.96\widehat{\mathrm{SE}}(\widehat{\beta})$ (where $\widehat{\mathrm{SE}}(\widehat{\beta})$ depends on the variance estimation method used). Coverage is defined by the proportion of times $\beta$ is included in the 95\% confidence interval of $\beta$ estimated from the model.
\end{enumerate}

\subsection{Software}\label{ss-software}

All simulations involved the use of \verb!R! 3.5.\cite{r} Sandwich variance estimation were computed with the \texttt{svyglm} function from the \texttt{survey} package (\cite{lumley_analysis_2004}).
The \texttt{boot} package  was used for boostrap sampling (\cite{boot_package}).
Graphics were generated using the \texttt{ggplot2} package (\cite{wickham_ggplot2elegantgraphics_2009}).

\subsection{Results}

\begin{figure}[!htb]
	\centerline{\includegraphics[width=15cm,height=15cm]{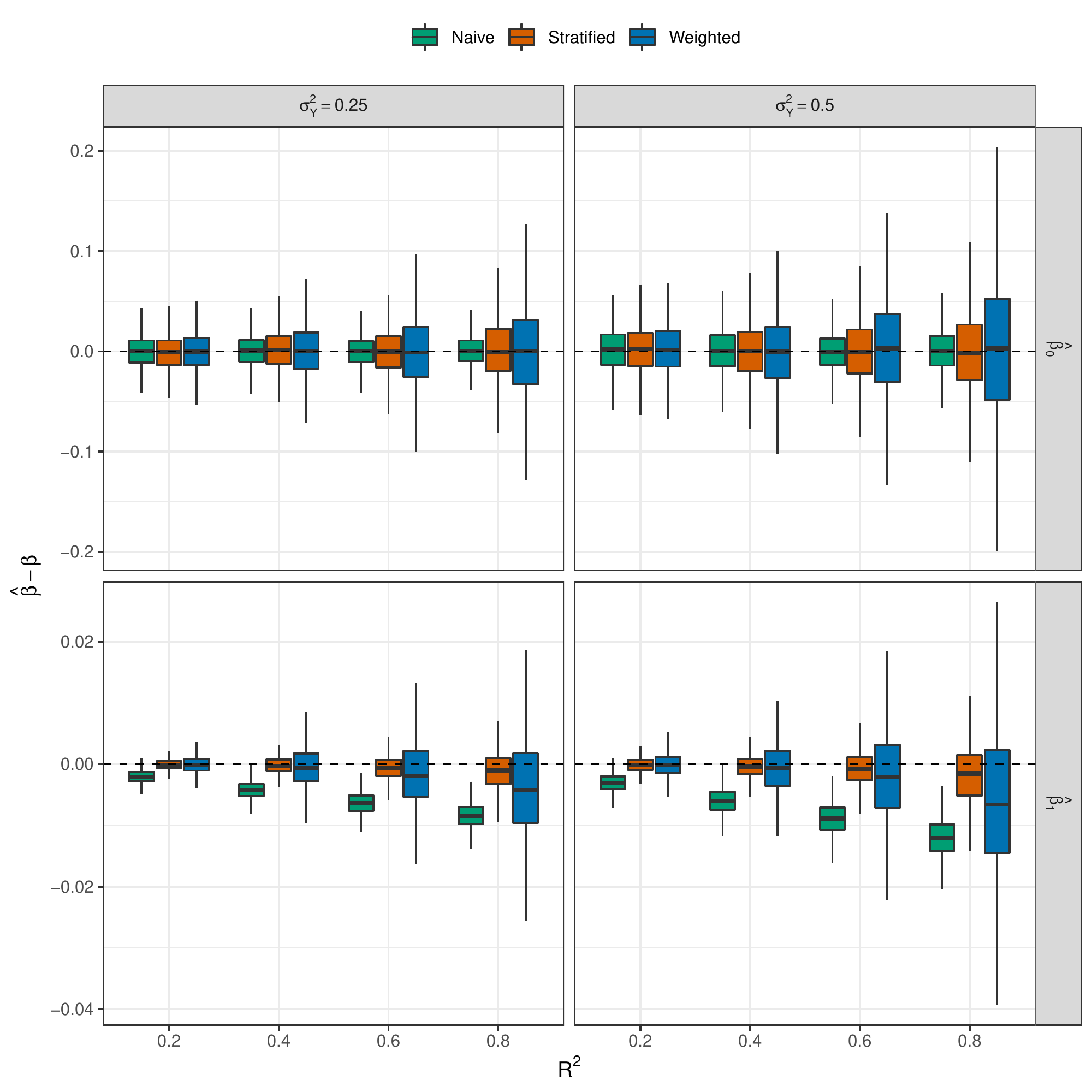}}
	%\centerline{\includegraphics[width=18cm]{figs/dag.png}}
	%\centerline{The figure was removed and submitted separately.}
	\caption{Boxplots of the difference between  $\widehat{\beta}$ and $\beta$ for different values of  $R^2 \in \{0.2,0.4,0.6,0.8\}$ and $\sigma_Y^2 \in \{0.25,0.5\}$, $\beta_0=0$, $\beta_1=1$ and $n =1000$.}
	\label{fig:bias}
\end{figure}

\begin{figure}[!htb]
	\centerline{\includegraphics[width=15cm,height=15cm]{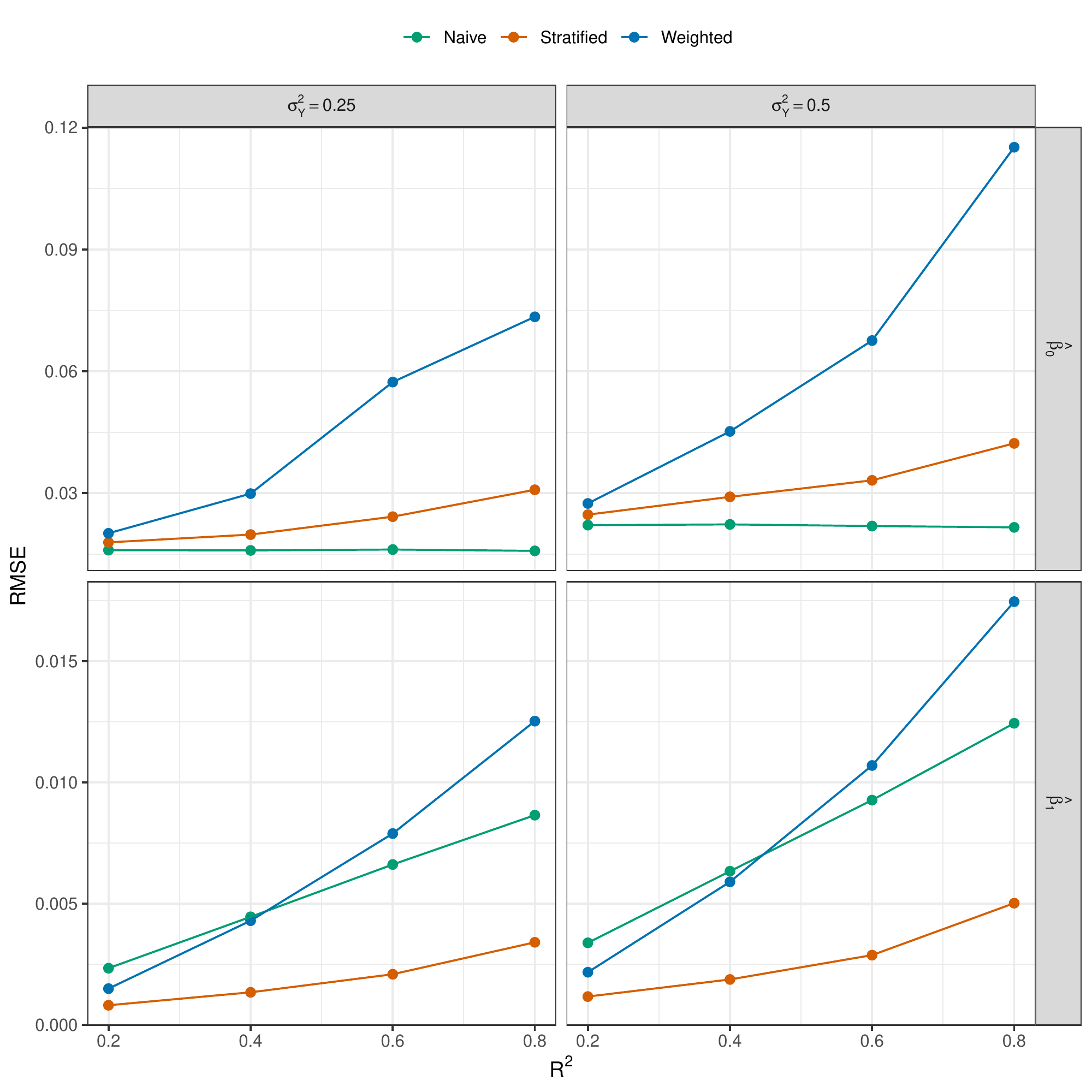}}
	\caption{Root mean square error (RMSE) of  $\widehat{\beta}$ for different values of  $R^2 \in \{0.2,0.4,0.6,0.8\}$ and $\sigma_Y^2 \in \{0.25,0.5\}$, $\beta_0=0$, $\beta_1=1$ and $n =1000$.}
	\label{fig:RMSE}
\end{figure}

For each of the two parameters of the dose-response function ($\beta_0$ and $\beta_1$), Figure \ref{fig:bias} displays the bias of the estimates for different values of $R^2$ and $\sigma_Y^2$. In this figure, the theoretical values of $\beta_0$ (the intercept coefficient of the dose-response function) and $\beta_1$ (the slope coefficient) are 0 and 1 respectively. Boxplots were plotted to allow the graphical assessment of the variability of the estimates. For the estimation of $\beta_1$ (lower panel), the performance of all methods was highly influenced by the value of $R^2$, with a (negative) bias which increased with increasing $R^2$. This may be explained by the fact that the more $R^2$ increases, the less the positivity assumption is respected. Also, the bias increases with the value of $\sigma_Y^2$. Overall, the stratification method gave the smallest bias, while the naive method gave the largest bias. The weighted method gave acceptable bias only for value of $R^2\leq 0.4$. All methods seemed to give approximately unbiased estimates of $\beta_0$ (upper panel), but the graphical evaluation of the bias is made difficult by the very high variability of the estimates. In fact, the same trends as previously reported for $\beta_1$ were observed for $\beta_0$ estimates, except that the bias was positive instead of negative. The variability of the estimates increased for all methods as $R^2$ and $\sigma_Y^2$ increased. The variability associated with the weighted estimator seemed much larger than with the stratified or the naive estimator, particularly for the estimation of $\beta_1$. The combination of a significant bias and a very high variability for large $R^2$ values led to the highest RMSE values being observed for the weighted method (Figure \ref{fig:RMSE}). On the contrary, the stratification method was associated with the lowest RMSE, regardless of the $R^2$ and $\sigma_Y^2$ values for the estimation of both $\beta_0$ and $\beta_1$.

\begin{figure}[!htb]
	\centerline{\includegraphics[width=15cm,height=18.75cm]{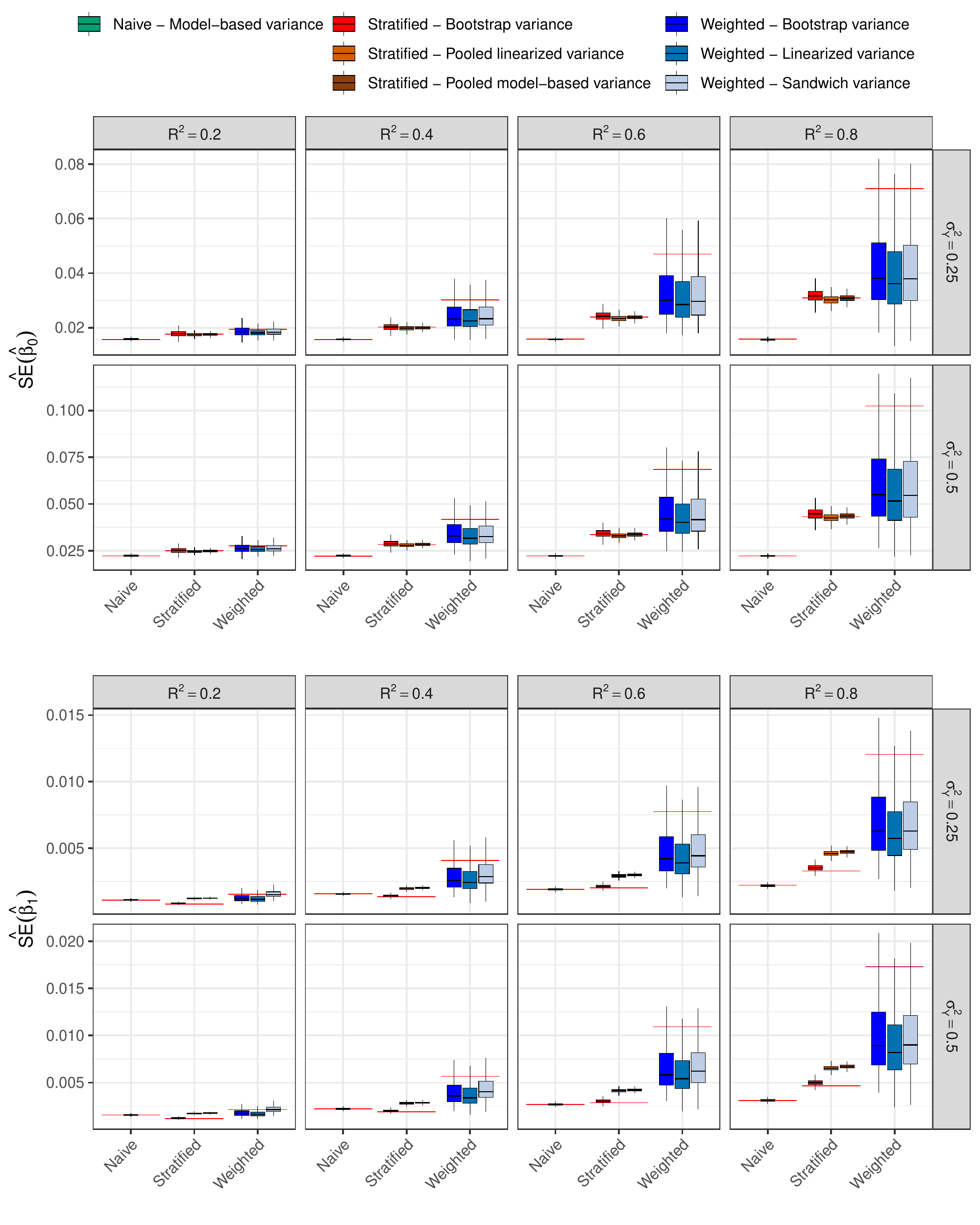}}
	\caption{Boxplots of the standard deviation estimates of  $\widehat{\beta}$ for different values of  $R^2 \in \{0.2,0.4,0.6,0.8\}$ and $\sigma_Y^2 \in \{0.25,0.5\}$, $\beta_0=0$, $\beta_1=1$ and $n =1000$. The red line corresponds to the Monte Carlo standard deviation estimate.}
	\label{fig:stdr}
\end{figure}

Figure \ref{fig:stdr} displays the boxplots of the standard errors of the dose-response function parameters ($\beta_0$ in the upper panel, $\beta_1$ in the lower panel) estimated with all methods listed in Table \ref{tab:methods}. The empirical Monte-Carlo standard deviation associated with each dose-response function parameter estimator in each evaluated scenario was indicated by a red horizontal line. In all scenarios, the highest empirical standard deviation were observed with the weighted estimator of the dose-response function parameters, especially for large values of $R^2$. The empirical standard deviation estimates associated with the stratified estimator of the dose-response function were higher than those associated with the naive estimator, especially for an $R^2$ value of 0.8. The value of $\sigma_Y^2$ had less effect than $R^2$ on empirical standard deviations. For the weighted estimator of the dose-response function, the standard errors estimated with boostrap, linearized and sandwich methods underestimated the empirical standard deviation. The variability of these standard error estimates was also very large, and this phenomenon increased with $R^2$. Among variance estimation methods associated with the stratified estimator of the dose-response function, the bootstrap (which take into account the GPS estimation step) produced the closest estimation of the empirical standard deviation of $\hat{\beta_1}$ coefficient. The two other estimators (pooled linearized and pooled model-based estimators) overestimated the standard deviation of the $\hat{\beta_1}$ coefficient. All variance methods associated with the stratified estimator produced reasonably good estimates of the variance of $\hat{\beta_0}$. Finally, the model-based variance estimator of the naive dose-response function estimator had good performance in all scenarios.

\begin{figure}[!htb]
	\centerline{\includegraphics[width=15cm,height=15cm]{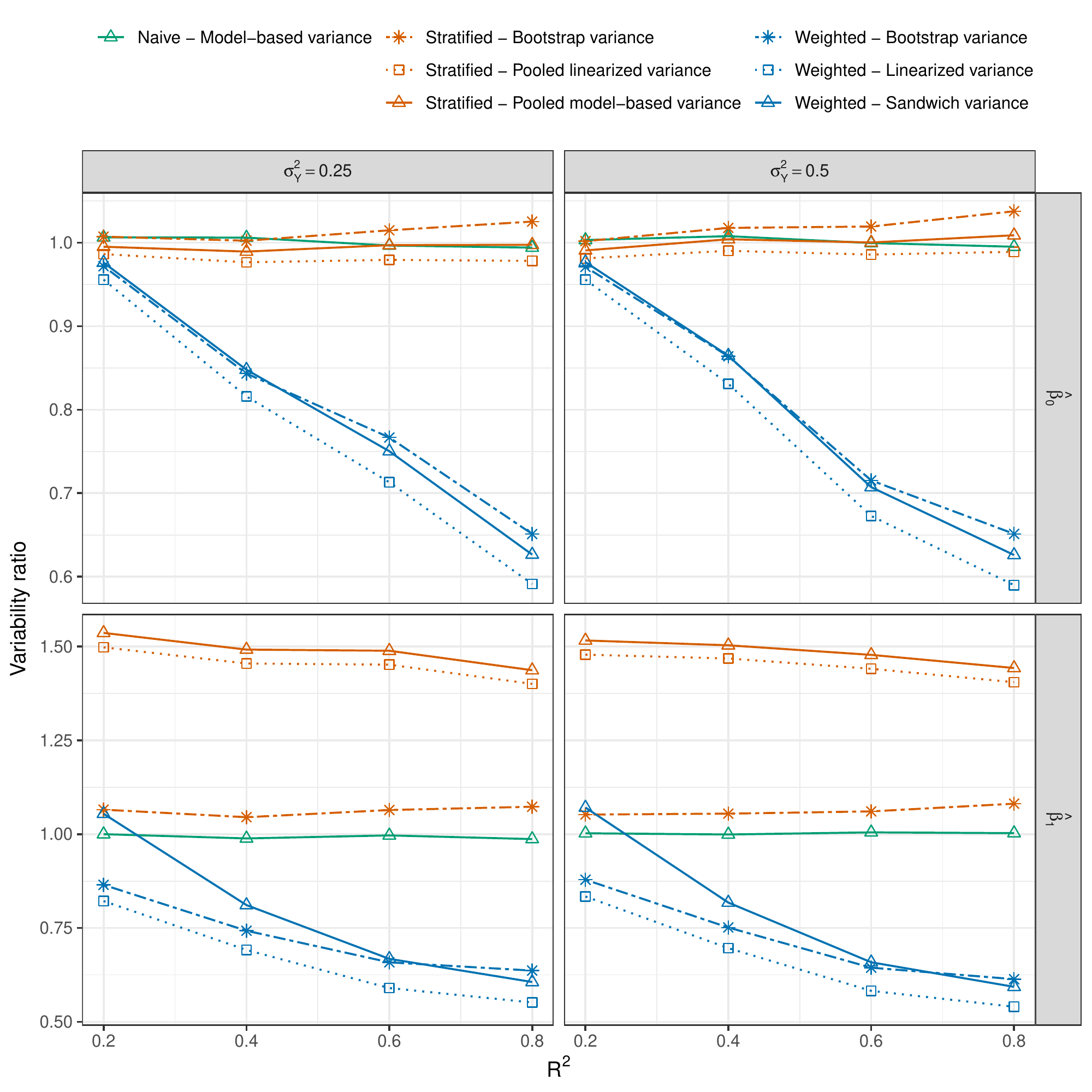}}
	\caption{Variability ratio for different values of $R^2 \in \{0.2,0.4,0.6,0.8\}$ and $\sigma_Y^2 \in \{0.25,0.5\}$, $\beta_0=0$, $\beta_1=1$ and $n =1000$.}
	\label{fig:var_ratio}
\end{figure}

The patterns of over or underestimation of the empirical standard deviation are more precisely observable in
Figure \ref{fig:var_ratio}, which illustrates the ratio of the average standard error and empirical standard deviation of the intercept (upper panel) and the slope (lower panel) estimated coefficients for different values of parameters $R^2$ and $\sigma_Y^2$. Overall, sandwich, bootstrap and linearized variance estimators of the weighted estimator of dose-response function resulted in similar values of variability ratio and were negatively biased, except for the sandwich variance estimator of $\beta_1$ with $R^2 = 0.2$. This underestimation of the empirical standard deviation increased with $R^2$, for the two coefficients of the dose-response function. Overall, the variance estimators of the stratified estimator performed well for the intercept parameter of the dose-response function, and tended to overestimate the empirical variance for the slope parameter. The pooled linearized variance method gave slightly lower variability ratios than the pooled model-based variance method. The boostrap variance method gave ratio values very close to 1 for the slope coefficient, whereas the two others estimators clearly led to an overestimation of the empirical variance. The performance of the three variance estimator of the stratified estimator stayed stable as the $R^2$ increased. Finally, the value of  $\sigma_Y^2$ did not affect substantially the previous description of the results for all estimators.

\begin{figure}[!htb]
	\centerline{\includegraphics[width=15cm,height=15cm]{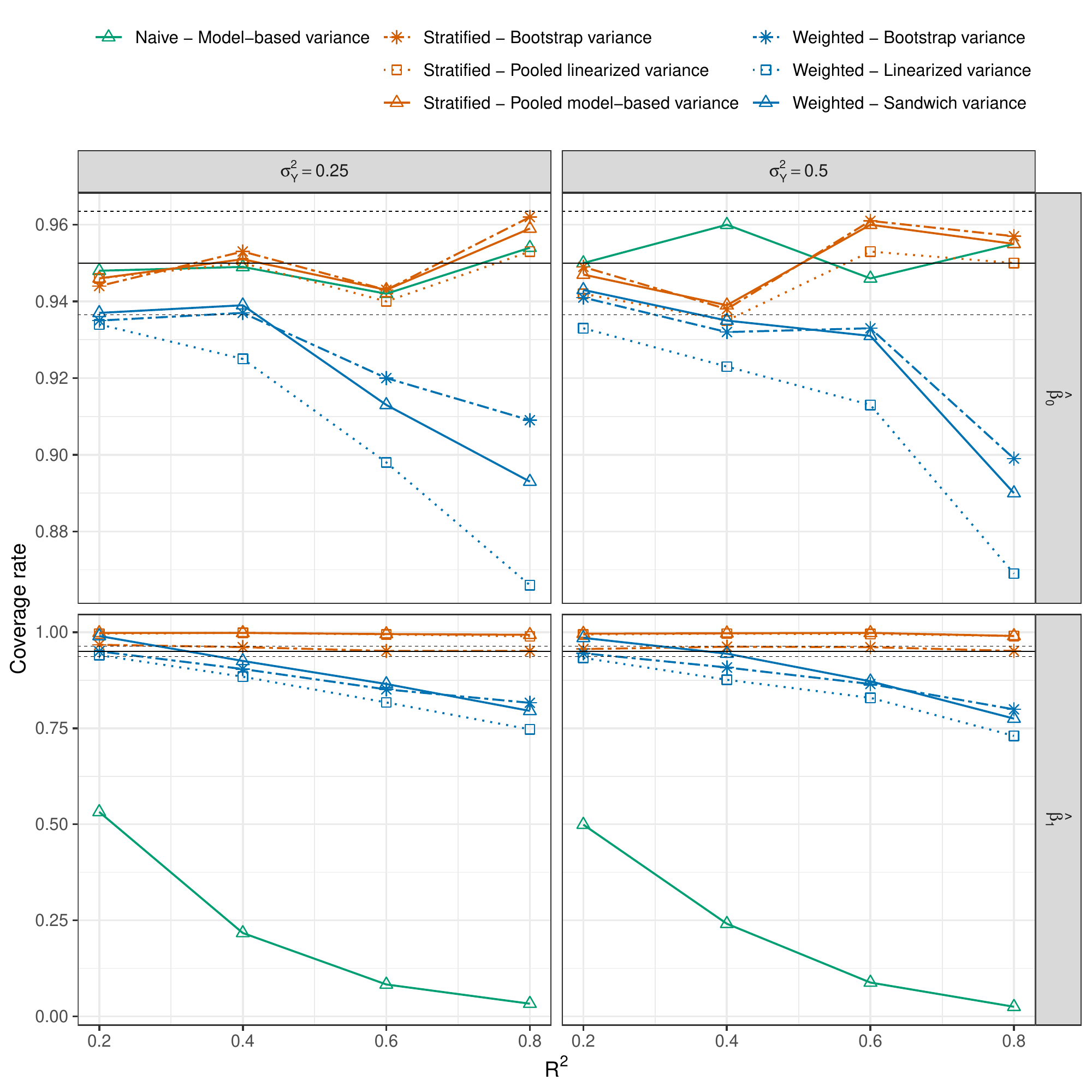}}
	\caption{Coverage rate for different values of  $R^2 \in \{0.2,0.4,0.6,0.8\}$ and $\sigma_Y^2 \in \{0.25,0.5\}$,  $\beta_0=0$, $\beta_1=1$ and $n =1000$.}
	\label{fig:Coverage}
\end{figure}

Finally, coverage rates for dose-response function estimates are reported on Figure \ref{fig:Coverage} for different values of parameters $R^2$ and $\sigma_Y^2$. Overall, results were consistent with those described previously. For the intercept coefficient of the dose-response function (upper panel), confidence intervals based on weighted estimator of the dose-response function were the most unconservative. Their coverage rate decreased as the $R^2$ increased, whereas the performance of other estimators performed well in all scenarios. Pooled model-based and pooled linearized variance estimator of the stratified estimator of the dose-response function were too conservative for the estimation of $\beta_1$, while boostrap-based method gave approximately correct coverage rates. Again, the performance of the weighted estimators was greatly influenced by $R^2$ values, with coverage rates deteriorating while $R^2$ increased.

Supplementary simulations showed that these different results were not affected by a change in the theoretical value of $\beta_1$ (see Supplementary materials, Section 1). The effect of the sample size $n$ had also been studied and showed that the bias and variability associated with the different estimators increased as the sample size decreased (see Supplementary materials, Section 2).

Finally, we also studied the effect of different number of strata for the stratified method on the different estimations  (see Supplementary materials, Section 3). The different estimations of $\beta_0$  were better for a few number of strata, while the different estimations for $\beta_1$  were better for a large number of strata. The variance estimation of $\widehat{\beta_0}$  and $\widehat{\beta_1}$ were better for a few number of strata.\\

\section{Real data application}\label{example}

The different dose-response function estimation methods and associated variance estimation methods have been applied on a real cohort extracted from the PreCARE study. PreCARE is a prospective multicenter cohort study of pregnant women aiming to examine the association between socioeconomic exposure and adverse maternal or neonatal outcomes (\cite{gonthier_association_2017}). It included all consecutive women registered to deliver or who delivered in 4 public teaching hospitals in northern Paris (France) between October 2010 and May 2012. Women were included at the beginning of their pregnancy during their first visit at 1 of the 4 facilities and were followed until hospital discharge after delivery. Overall, 10,419 women and their newborns were included. The objective of this analysis was to study the relationship between pre-pregnancy maternal body mass index (BMI) on newborn birth weight. This analysis was based on the 8,775 women for whom information about BMI, birth weight and confounding factors included in the propensity score model was available. The list of co-variables included in the propensity score model was maternal age, parity, history of pre-eclampsia, history of preterm delivery (ie, before 37 weeks' gestational age), the presence of a social deprivation (social deprivation index $\geq 1$) (\cite{kantor_socioeconomic_2017}), and maternal birth place (France vs other). The $R^2$ value associated with the propensity model was 0.05.

All the statistical methods described in Section \ref{ss-methods} were applied. The estimated parameters of the dose-response function are reported in Table \ref{tab:real}. In this Table, $\hat{\beta}_0$ represents the estimated mean birth weights (in grams) when maternal BMI is equal to 10, and $\hat{\beta}_1$ represents the increment of the estimated mean birth weight when maternal BMI increases of 1 unit. All $\hat{\beta}_1$ values had qualitatively similar values indicating the positive association between maternal BMI and birth weight. Of note, the graphical inspection of the relationship between maternal BMI and birth weight may suggest a 'plateau' effect for the highest (and rarest) BMI values (\cite{froslie_categorisation_2010}). For the sake of simplicity, this eventual deviation from the linearity hypothesis of the relationship between maternal BMI and birth weight has been neglected.

As in the simulation study, standard errors associated with the weighted dose-response function estimation method were higher than those associated with the stratified estimation method, indicating a higher variability of maternal BMI effect estimate. Among variance estimation methods associated with the weighted estimator, the methods which take into account the GPS estimation step (linearized and bootstrap) produce lower standard errors than the sandwich method. The same was observed among the variance estimation methods associated with the stratified estimator: bootstrap standard error was lower than the standard errors estimated with the two other methods which do not take into account the GPS estimation step. Overall, these results were consistent with the results observed in the simulation study.

\begin{table}
	\caption{Estimated dose-response function in the case study.}
	\label{tab:real}
	\centering
	
	\begin{tabular}{lrr}
		%\toprule
        \hline
		& \multicolumn{1}{c}{$\hat{\beta}_0$} & \multicolumn{1}{c}{$\hat{\beta}_1$} \\
		Method & \multicolumn{1}{c}{$(\widehat{\mathrm{SE}}(\widehat{\beta}_0))$} & \multicolumn{1}{c}{$(\widehat{\mathrm{SE}}(\widehat{\beta}_1))$} \\
		%\midrule
        \hline
		Naive                                                  & 3065.41 & 12.78  \\
		\hspace*{0.5cm}Model-based standard error              & (20.62) & (1.42) \\
		Weighted                                               & 3116.06 & 8.98   \\
		\hspace*{0.5cm}Sandwich standard error                 & (32.02) & (2.31) \\
		\hspace*{0.5cm}Linearized standard error               & (29.18) & (2.10) \\
		\hspace*{0.5cm}Bootstrap standard error                & (31.17) & (2.25) \\
		Stratified                                             & 3081.81 & 11.61  \\
		\hspace*{0.5cm}Pooled model-based standard error       & (20.09) & (1.33) \\
		\hspace*{0.5cm}Pooled linearized standard error        & (21.17) & (1.45) \\
		\hspace*{0.5cm}Bootstrap standard error                & (17.32) & (1.13) \\
		%\bottomrule
        \hline
	\end{tabular}
	
	$\widehat{\mathrm{SE}}(\hat{\beta}_p)$: estimated standard error
\end{table}

\section{Conclusions and perspectives}\label{conclusion}

GPS-based methods have been proposed as a generalization of the propensity score framework for assessing the marginal effect of a quantitative exposure on an outcome of interest, through the estimation of a dose-response function. This research focuses on the variance estimation of the dose-response function parameters, in the case of continuous outcome. We have considered different dose-response function estimation methods and different variance estimation for each methods.

Experimental tests on simulated databases show that the stratification method gives the best estimation of the parameters of the dose-response function, and the boostrap method gives the best estimation of the associated variance. The pooled variance estimators (using linearized or maximum likelihood model-based estimator) of the stratified estimator overestimate the variance, resulting in estimated 95\% confidence intervals whose empirical coverage rates are substantially higher than the nominal level. This phenomenon is related to what was already reported in the case of binary exposure: the use of a variance estimator that does not account for the fact that the propensity score is estimated rather than known with certainty leads to an overestimation of the variability of the estimate of treatment effect.

At the beginning of this research project, our main objective was to develop of a closed-form estimator of the variance of the coefficients of the dose-response function estimated using GPS-weighting, taking into account the weights estimation step. But after the evaluation of the performance of GPS-weighting and of the proposed variance estimator, as well as the performance of the bootstrap estimator (already used in \cite{zhang2016causal}), we had to admit that our enthusiasm about GPS-weighting was dampened. This study shows that GPS-weighting adds up three important issues: a greater bias than the stratified method as the $R^2$ increases, a high variability of the estimates, and the failure of different variance estimators to correctly capture this variability, even though similar approaches have been successfully used in the context of propensity score weighting for binary exposure (\cite{williamson_variancereductionrandomised_2014, austin_varianceestimationwhen_2016, hajage2018closed}). Even if the bias observed in simulations was relatively limited and became really significant for large values of $R^2$, the high variability of estimates led to RMSE values equal to or greater than those observed without any adjustment. Moreover, the underestimation of the variance with all variance estimators led to coverage rates well below the nominal level. These shortcomings lead us to not recommend the use of GPS-weighting for assessing the effect of a quantitative exposure in observational studies, and to prefer more efficient alternative methods like the stratification method that was also evaluated in this study, or covariate adjustment using the GPS which also seems to provide more accurate estimates than GPS-weighting, as shown by \cite{austin2018assessing} for a binary outcome.

Given the performance of the stratified estimator combined with bootstrap variance estimation, a future research may focus on the development of a closed-form variance estimator which takes into account the fact that strata boundaries are estimated rather than known. This could be particularly useful for the analysis of very large datasets (such as healthcare-administrative databases), for which the repeated calculations required for the bootstrap methods could be an issue (see Supplementary materials, Figure 16, for a comparison of execution times recorded with each variance estimation method). Another topic of research could seek to improve the performance of the weighted estimator. As suggested by Austin, 'large value of $R^2$ results in some subjects having large weights, resulting in estimates with high variability' (\cite{austin2018assessing}). The use of the marginal density of the quantitative exposure to stabilize the estimated weights was already shown to significantly improve the performance of GPS-weighting compared to the use of unstabilized weights (\cite{zhao2020propensity}). Nevertheless, our study shows that this stabilization fails to make the method competitive with the simple alternative that is stratification. Perhaps a different choice for the numerator of the GPS-weights could help to reduce even more the unstability and improve the overall performance. Another research perspective would be to study more complex dose-response functions. Indeed, our research was deliberately limited to the study of simple linear response models with only two parameters (the intercept and the slope), because the main objective of our study was not to compare different approaches for estimating a complex function, but to study the ability of various variance estimators to capture the variability of the estimates. While the inclusion of polynomial terms in the weighted model does not raise any particular difficulty, studying more complex models, including smooth coefficients or non-parametric modelization of the dose response function (\cite{zhao2020propensity}) would be interesting in order to get closer to real clinical situations in whom dose-response functions are not always linear. But the development and evaluation of variance estimators (including the comparison to the empirical variance estimation) adapted to these situations is not simple, and was beyond the scope of this work.

%\backmatter

\section*{Acknowledgments}
The French Ministry of Health funded the PreCARE study. The authors thank Elie Azria for his permission to use the data from the PreCARE cohort, and his insightful comments on the analysis of this case study.

\section*{Data availability statement}

The data from the case study are available on request from the corresponding author upon reasonable request. The data are not publicly available due to privacy or ethical restrictions.

\section*{Conflict of interest}

The authors declare no potential conflict of interests.

\section*{Supporting information}

Additional simulation results may be found in the online version of this article at the publisher’s
web site.

\clearpage

\section*{Figures}

\clearpage

\appendix

\section{Variance estimator for the weighted dose-response function estimator} \label{appen1}

We write
\begin{eqnarray} \label{lin:eq1}
H_n (\hat{\gamma},\hat{\beta})-H_n (\gamma,\beta) & = & \frac{1}{n} \sum_{i=1}^n w_i(\hat{\gamma}) \{H_i(\hat{\beta})-H_i(\beta)\} \\
& + & \frac{1}{n} \sum_{i=1}^n \left\{w_i(\hat{\gamma})-w_i(\gamma)\right\} H_i(\beta). \nonumber
\end{eqnarray}
We first consider the first term in the right-hand side of (\ref{lin:eq1}), denoted as $\Delta_1$. Making use of a first-order Taylor expansion, we obtain
\begin{eqnarray} \label{lin:eq2}
\Delta_1 & = & \frac{1}{n} \sum_{i=1}^n w_i(\hat{\gamma}) \{\nabla H_i(\beta)^{\top}(\hat{\beta}-\beta)+o_p(n^{-0.5})\} \nonumber \\
& = & \left(\frac{1}{n} \sum_{i=1}^n w_i(\gamma) \nabla H_i(\beta) \right) (\hat{\beta}-\beta) + o_p(n^{-0.5}) \nonumber \\
& = & -A (\hat{\beta}-\beta) + o_p(n^{-0.5}),
\end{eqnarray}
where
\begin{eqnarray*} \label{lin:eq3}
	A & = & E[-w_i(\gamma) \nabla H_i(\beta)].
\end{eqnarray*}
In view of the system of estimating equations (\ref{ee:beta}), we have
\begin{eqnarray*}
	\nabla H_i(\beta) & = & - \tilde{T}_i \tilde{T}_i^{\top},
\end{eqnarray*}
which leads to
\begin{eqnarray} \label{lin:eq5}
A & = & E[w_i(\gamma) \tilde{T}_i \tilde{T}_i^{\top}].
\end{eqnarray}

We now consider the second term in the right-hand side of (\ref{lin:eq1}), denoted as $\Delta_2$. Making use of a first-order Taylor expansion, we obtain
\begin{eqnarray} \label{lin:eq4}
\Delta_2 & = & \frac{1}{n} \sum_{i=1}^n \left\{\nabla w_i(\gamma))^{\top} (\hat{\gamma}-\gamma) + o_p(n^{-0.5})\right\} H_i(\beta) \nonumber \\
& = & \left(\frac{1}{n} \sum_{i=1}^n H_i(\beta) \nabla w_i(\gamma)^{\top} \right) (\hat{\gamma}-\gamma) + o_p(n^{-0.5}) \nonumber \\
& = & B (\hat{\gamma}-\gamma) + o_p(n^{-0.5}),
\end{eqnarray}
where
\begin{eqnarray}  \label{lin:eq6}
B & = & E[H_i(\beta) \nabla w_i(\gamma)^{\top}].
\end{eqnarray}
Making use of equation (\ref{gps:weight}), we obtain after some algebra
\begin{eqnarray} \label{eq:nabla:wi}
\nabla w_i(\gamma) & = & w_i(\gamma)
\left(
\begin{array}{l}
\frac{T_i-\mu_T}{\sigma_T^2} \\
\frac{1}{2\sigma_T^2} \left\{\frac{(T_i-\mu_T)^2}{\sigma_T^2}-1\right\} \\
-\tilde{Z}_i \frac{(T_i-\tilde{Z}_i^{\top} \alpha)}{\sigma^2} \\
-\frac{1}{2\sigma^2} \left\{\frac{(T_i-\tilde{Z}_i^{\top} \alpha)^2}{\sigma^2}-1\right\}
\end{array}
\right).
\end{eqnarray}

Since $\gamma$ is estimated by solving the estimating equation (\ref{Fn:gamma}), we also have
\begin{eqnarray*}
	F_n(\hat{\gamma})-F_n(\gamma) & = & \nabla F_n(\gamma)^{\top} \left\{\hat{\gamma}-\gamma \right\}+o_p(n^{-0.5}),
\end{eqnarray*}
and since $F_n(\hat{\gamma})=0$, this leads to
\begin{eqnarray}
\hat{\gamma}-\gamma & = & -\left\{\nabla F_n(\gamma)\right\}^{-1} \times F_n(\gamma) + o_p(n^{-0.5}) \nonumber \\
& = & C^{-1} \times F_n(\gamma) + o_p(n^{-0.5})
\end{eqnarray}
where
\begin{eqnarray*}
	C & = & E[-\nabla F_n(\gamma)].
\end{eqnarray*}
From the definition of $F_n$ given in equation (\ref{Fn:gamma}), we obtain after some algebra
\begin{eqnarray*}
	C & = & \left(
	\begin{array}{cccc}
		1 & 0 & 0_{1K} & 0 \\
		0 & \frac{n-1}{n} & 0_{1K} & 0 \\
		0_{K1} & 0_{K1} & E(\tilde{Z}_i \tilde{Z}_i^{\top}) & 0_{K1} \\
		0 & 0 & 0 & \frac{n-p-1}{n}
	\end{array}
	\right).
\end{eqnarray*}

\bigskip

By gathering equations (\ref{lin:eq1}), (\ref{lin:eq2}), (\ref{lin:eq4}) and (\ref{lin:eq6}), and since $H_n (\hat{\beta})=0$, we obtain
\begin{eqnarray*}
	-H_n (\gamma,\beta) & = & -A (\hat{\beta}-\beta) + B C^{-1} F_n(\gamma) + o_p(n^{-0.5}),
\end{eqnarray*}
which finally gives
\begin{eqnarray}
\hat{\beta}-\beta & = & \frac{1}{n} \sum_{i=1}^n I_{1i}+o_p(n^{-0.5}), \\
\textrm{where } I_{1i} & = & A^{-1} \left\{w_i(\gamma)H_i(\beta)+BC^{-1}F_i(\gamma) \right\}.
\end{eqnarray}
The variable $I_{1i}$ is the theoretical linearized variable of $\hat{\beta}$. It involves unknown parameters, which need to be estimated for variance estimation. This leads to the estimated linearized variable $\hat{I}_{1i}$ given in equation (\ref{lin:weight:1}).

\newpage

\section{Variance estimator for the stratified dose-response function estimator} \label{appen2}

Recall that the intermediary estimator $\hat{\beta}_{\ell},~\ell=1,\ldots,L,$ is obtained by solving the estimating equation (\ref{strat:est:eq:1}). We have
\begin{eqnarray} \label{app2:eq1}
\Phi(\widehat{\beta}_{\ell})-\Phi(\beta_{\ell}) & = & -\Phi(\beta_{\ell}) \nonumber \\
& \simeq & E\left\{ \nabla \Phi(\beta_{\ell})\right\} \times \{\widehat{\beta}_{\ell}-\beta_\ell\}.
\end{eqnarray}
This leads to
\begin{eqnarray} \label{app2:eq2}
\widehat{\beta}_{\ell}-\beta_\ell & \simeq & - E\left\{ \nabla \Phi(\beta_{\ell})\right\}^{-1} \times \Phi(\beta_{\ell}) \nonumber \\
& = & \frac{1}{\sigma_{\ell,T}^2} \left(
\begin{array}{cc}
\sigma_{\ell,T}^2+m_{\ell,T}^2 & -m_{\ell,T} \\
-m_{\ell,T} & 1
\end{array}
\right) \times \frac{1}{n_\ell} \sum_{i \in S_\ell} \left(
\begin{array}{l}
Y_i - \tilde{T}_i^{\top} \beta_{\ell} \\
T_i(Y_i -\tilde{T}_i^{\top} \beta_{\ell})
\end{array}
\right) \nonumber \\
& = &  \frac{1}{n_\ell} \sum_{i \in S_\ell} I_{2l,i}
\end{eqnarray}
with
\begin{eqnarray} \label{app2:eq3}
I_{2l,i} & = & \frac{1}{\sigma_{\ell,T}^2}
\left(
\begin{array}{cc}
\sigma_{\ell,T}^2+m_{\ell,T}^2 & -m_{\ell,T} \\
-m_{\ell,T} & 1
\end{array}
\right) \times
\left(
\begin{array}{l}
Y_i - \tilde{T}_i^{\top} \beta_{\ell} \\
T_i(Y_i - \tilde{T}_i^{\top} \beta_{\ell})
\end{array}
\right),
\end{eqnarray}
and where $m_{\ell,T}=E\left\{T_i \mathbf{1}_{\left\{i \in S_\ell\right\}}\right\}$ and $\sigma_{\ell,T}^2=V\left\{T_i \mathbf{1}_{\left\{i \in S_\ell\right\}}\right\}$. \\

The variable $I_{2l,i}$ is the theoretical linearized variable of $\hat{\beta}_{\ell}$. Replacing the unknown parameters by suitable estimators leads to the estimated linearized variable $\hat{I}_{2l,i}$ given in (\ref{strat:est:eq:3}).

\newpage

\section{R code for the different variance estimators} \label{appen3}

\begin{singlespace}
\begin{verbatim}
######################################################################
# This code is provided for illustrative purposes only and comes with
# absolutely NO WARRANTY.
######################################################################
library(survey)
library(boot)

######################################################################
# Weight estimation
######################################################################

# Fit the propensity model. Trt is the exposure, Z1 to Z10 are the covariates
modT <- lm(Trt ~ Z1 + Z2 + Z3 + Z4 + Z5 + Z6 + Z7 + Z8 + Z9 + Z10, data = data)

# Linear predictor
data$m <- m <- modT$fitted

# Computation of the estimated weights
n <- nrow(data)
s <- sqrt(sum(modT$residuals^2)/(n-length(modT$coef)))
wd <- dnorm(data$Trt, m, s)
mu <- mean(data$Trt)
su <- sd(data$Trt)
wn <- dnorm(data$Trt, mu, su)

data$w <- w <- wn/wd

######################################################################
# Weighted estimator - sandwich standard error
######################################################################

mod <- svyglm(Y ~ Trt, design = svydesign(id = ~1, weights = ~ w, data = data),
       family = gaussian)
summary(mod)

######################################################################
# Weighted estimator - linearized standard error
######################################################################

coefs.ipw <- mod$coefficients

variables <- names(data)[grep("^Z", names(data))]
Z <- as.matrix(data[, variables])
Ztilde <- cbind(1, Z)
dw <- w*cbind(
  (data$Trt - mu)/(su^2),
  (((data$Trt - mu)/su)^2 - 1)/(2*su^2),
  -as.vector((data$Trt - data$m)/(s^2))*Ztilde,
  -(((data$Trt - data$m)/s)^2 - 1)/(2*s^2)
)

Ttilde <- cbind(1, data$Trt)
tmp <- cbind(Ttilde, data$Trt, data$Trt^2)
A <- matrix(colMeans(tmp * w), 2, 2)
sA <- solve(A)

H <- Ttilde*as.vector((data$Y - mod$fitted.values))

F <- cbind(
  data$Trt - mu,
  (data$Trt - mu)^2 - ((n-1)/n)*(su^2),
  as.vector((data$Trt - data$m))*Ztilde,
  ((data$Trt - data$m)^2) - ((n-length(modT$coef))/n)*(s^2)
)

B <- crossprod(H, dw)/n

mZZ <- crossprod(Ztilde, Ztilde)/n
C <- diag(length(variables) + 1 + 3)
C[2, 2] <- (n-1)/n
C[3:(length(variables)+3), 3:(length(variables)+3)] <- mZZ
C[nrow(C), ncol(C)] <- ((n-length(modT$coef))/n)
sC <- solve(C)

I <- t(sA%*%t((w*H + t(B %*% sC%*%t(F)))))

sds.ipw.lin <- sqrt(apply(I, 2, var)/n)
names(sds.ipw.lin) <- names(coefs.ipw)

print(coefs.ipw)
print(sds.ipw.lin)

######################################################################
# Weighted estimator - bootstrap standard error
######################################################################
f.boot.ipw <- function(data, i) {
  df <- data[i, ]
  modT <- lm(Trt ~ Z1 + Z2 + Z3 + Z4 + Z5 + Z6 + Z7 + Z8 + Z9 + Z10, data = df)
  m <- modT$fitted

  n <- nrow(df)
  s <- sqrt(sum(modT$residuals^2)/(n-length(modT$coef)))
  wd <- dnorm(df$Trt, m, s)
  mu <- mean(df$Trt)
  su <- sd(df$Trt)
  wn <- dnorm(df$Trt, mu, su)

  df$w <- wn/wd

  lm.wfit(cbind(rep(1, nrow(df)), df$Trt), df$Y, df$w)$coef
}

rcoefs <- boot(data, f.boot.ipw, R = 200)$t
sds.ipw.boot <- apply(rcoefs, 2, sd)
names(sds.ipw.boot) <- names(coefs.ipw)
print(sds.ipw.boot)

######################################################################
# Stratified estimator - Pooled model-based standard error
######################################################################
cl <- 10 # number of strata
data$Tcl <- cut(data$m, breaks = quantile(data$m, probs = seq(0, 1, 1/cl)),
            include.lowest = TRUE)
W1 <- apply(data.frame(levels(data$Tcl)), MARGIN = 1, function(x) {
  data2 <-  subset(data, data$Tcl == x)
  nk <- nrow(data2)
  pk <- nk/n
  mod <- glm(Y ~ Trt, data = data2, family = gaussian)
  coefs <- mod$coef
  sds <- (summary(mod)$coefficients[,2])^2
  return(c(pk*coefs, pk^2*sds))
})
coefs.strat <- apply(W1[1:2,], MARGIN = 1, sum)
sds.strat.pool1 = sqrt(apply(W1[3:4,],MARGIN = 1, sum))
print(coefs.strat)
print(sds.strat.pool1)

######################################################################
# Stratified estimator - Pooled linearized standard error
######################################################################
W2 <- apply(data.frame(levels(data$Tcl)), MARGIN = 1, function(x) {
  data2 <- subset(data,data$Tcl == x)
  nk <- dim(data2)[1]
  pk <- nk/n
    mod <- glm(Y ~ Trt, data = data2, family = gaussian)
    coefs <- mod$coef
    mhat <- mean(data2$Trt)
    shat <- var(data2$Trt)
    uhat <- rep((1/shat), nk)*as.vector(rep(shat+mhat^2,nk) -
            mhat*data2$Trt)*as.vector(data2$Y-coefs[1]-coefs[2]*data2$Trt)
    uhat2 <- rep((1/shat), nk)*as.vector(data2$Trt-mhat)
             *as.vector(data2$Y-coefs[1]-coefs[2]*data2$Trt)
    ubar <- mean(uhat)
    ubar2 <- mean(uhat2)
    sds1 <- 1/(nk*(nk-1))*sum((uhat-ubar)^2)
    sds2 <- 1/(nk*(nk-1))*sum((uhat2-ubar2)^2)
    return(c(pk*coefs, pk^2*sds1, pk^2*sds2))
})
sds.strat.pool2 = sqrt(apply(W2[3:4,], MARGIN = 1, sum))
print(sds.strat.pool2)

######################################################################
# Stratified estimator - Bootstrap standard error
######################################################################
f.boot.strat <- function(data, i) {
  df <- data[i, ]
  modT <- lm(Trt ~ Z1 + Z2 + Z3 + Z4 + Z5 + Z6 + Z7 + Z8 + Z9 + Z10, data = df)
  df$m <- modT$fitted
  df$Tcl <- cut(df$m, breaks = quantile(df$m, probs = seq(0, 1, 1/cl)),
            include.lowest = TRUE)

  W <- apply(data.frame(levels(df$Tcl)), MARGIN = 1, function(x) {
    df2 <-  subset(df, df$Tcl == x)
    nk <- nrow(df2)
    pk <- nk/n
    coefs <- lm.fit(cbind(rep(1, nrow(df)), df$Trt), df$Y)$coef
    return(c(pk*coefs))
  })
  apply(W[1:2,], MARGIN = 1, sum)
}

rcoefs <- boot(data, f.boot.strat, R = 200)$t
sds.strat.boot <- apply(rcoefs, 2, sd)
names(sds.strat.boot) <- names(coefs.strat)
print(sds.strat.boot)
\end{verbatim}
\end{singlespace}

\newpage

\clearpage

%\bibliography{wileyNJD-AMA}%

\end{document}